**Anomaly in coupled parametric resonance**

Adarsh Ganesan, Cuong Do and Ashwin Seshia

**We present experimental observations of an anomaly in the coupled response of auto-parametrically excited microelectromechanical beams. When one of the two coupled beams is driven at elevated amplitudes, the excitation of dominant and recessive modes is observed in the driven and non-driven beams respectively. This anomalous nature of auto-parametric excitation has been unexplored by both theory and experiments and falls outside the scope of the conventional description of parametric resonance.**

Experiments on coupled micro- and nano-mechanical structures have enabled the realization of multiple degree-of-freedom oscillators [1-4] and provided physical insight into collective and emergent response in such systems with greater control of system parameters than for equivalent experiments conducted at a macroscopic length scale. For example, studies concerning vibration mode localization have been performed in such systems exploring the effects of asymmetry and coupling strength with a greater degree of experimental control on the associated system parameters on testbeds that provide tight control of manufacturing tolerances and scalability [5,6]. The onset of parametric resonance in such systems is also facilitated by low damping conditions and high natural frequencies and the precise frequency control have enabled several practical studies of the associated phenomenon [7,8]. While many studies of direct parametric excitation and auto-parametric excitation in individual resonators have been reported, systematic experimental observations of parametric excitation in coupled structures have not been reported to date. Furthermore, theoretical studies [9-10] have also predicted interesting features in such coupled systems. This letter presents observations of anomalous behaviour in two coupled beam resonators that does not align with conventional models. Here, under asymmetric drive conditions, the observed auto-parametrically excited collective modes demonstrate features distinct from the case of symmetric drive conditions.

The experimental testbed is a microfabricated device consisting of two free-free beam structures of dimensions 1100 $\mu m$ X 350 $\mu m$ X 11 $\mu m$ coupled by a 20 $\mu m$ X 2 $\mu m$ X 11 $\mu m$ beam. 0.5 $\mu m$ thick AlN and 1 $\mu m$ thick Al layers deposited on the Si device layer allows for piezoelectric excitation of the device (see Supplemental Material). For preserving the mechanical symmetry of coupled structures, AlN and Al layers are patterned similarly on both the free-free beam devices. This also allows for independent actuation of both structures. The electrical signal generated from a waveform generator (Agilent 335ARB1U) is fed to the device wire-bonded within a Leaded Chip Carrier to the patterned 1 $\mu m$ thick Al electrodes. The response of the device is monitored using a Laser Doppler Vibrometer (LDV).

To demonstrate parametric excitation, a length extensional mode of the device at 3.86 $MHz$ is considered. An auto-parametric excitation of a secondary mode results when the device is actuated at sufficiently high amplitudes [11-13]. Here, an eigen-mode located at half of the drive frequency is parametrically excited by the driven mode and a parametric gain is associated with the sub-harmonic excitation. For the drive frequency 3.86 $MHz$, an auto-parametrically excited mode at 1.93 $MHz$ is observed. The eigenmodes of the device around 1.93 $MHz$ are predicted through the modal analysis in COMSOL Multiphysics. The two eigenmodes located at this frequency represent the dominant and recessive forms of an out-of-plane flexure mode and their mode shapes are presented in the figures 1C1 and 1C2 respectively.

When the two beams of the device are driven at the frequency 3.86 $MHz$ and power level 5 $dBm$, the dominant mode is parametrically excited on both the beams (Figure 2A). This result agrees with the model of strongly coupled parametric resonance. Under asymmetric drive conditions, the results presented in the figures 2B and 2C indicate that the dominant and recessive modes get simultaneously excited on the driven and non-driven beams respectively. However, a strongly coupled parametric resonance, irrespective of the symmetry of drive, should always lead to dominant mode excitation. The presence of recessive mode, as opposed to the dominant mode, on the non-driven beam is completely surprising. This can be explained by the limited degree of coupling between the beams as compared to a composite structure. Under this intermediate coupling level, a previously undetected pathway is present which leads to recessive mode excitation in the non-driven beam.

To systematically study this anomalous nature of coupled parametric resonance, drive level dependent experiments were carried out at the drive frequency 3.86 $MHz$. Figures 3A and 3D show the drive level dependence of motion at ω and ω/2 respectively when only beam A is driven. Until the drive power level reaches 0 $dBm$, the motion of the beams A and B at ω increases linearly. However, when the parametric threshold is reached, a sudden increase in the motion of beams A and B at ω/2 is observed (Figure 3D). As observed in the figure 2B, the motion of beams A and B at ω/2 correspond to dominant and recessive mechanical modes. Corresponding to the parametric excitation, the motion of beam A at ω is also increased (Figure 3A) through the quadratic nonlinearity of dominant mode. However, the motion of beam B at ω is saturated (Figure 3A) which can be explained by the absence of quadratic nonlinearity in the recessive mode. Figures 3B and 3E show the drive level dependence of motion at ω and ω/2 respectively when the beam B is only driven. Here, the resonance conditions are swapped and similar characteristics are produced. Figures 3C and 3F show the drive level dependence of motion at $\omega$ and $\omega/2$ respectively when both the beams are driven at equal drive levels. Here, the parametric excitation of dominant mode is established on both the beams. Due to the quadratic nonlinearity, the increase in the motion at $\omega$ is also observed soon after the parametric excitation.

The figures 2B and 2C show the parametric excitation of recessive mode in the non-driven beam and figure 2A shows the parametric excitation of dominant mode on both the driven beams. However, at low enough drive levels i.e. below parametric threshold in one of the beams, a recessive mode can only be parametric excited in that beam (Figure 4R1). Upon crossing the threshold, the excitation of dominant mode takes place (Figure 4R2). During this process, a concomitant increase in the displacement amplitude is also observed (Figure 4A).

This paper reports an anomaly in coupled parametric resonance. The relevance of this anomaly can be any coupled physical systems undergoing parametric instability. Also, the demonstration of this previously unexplored pathway in coupled parametric resonance can open up new opportunities in the design of parametric micromechanical resonators for the applications in Atomic Force Microscopy (AFM) [14], energy harvesters [15] and consumer devices such as MEMS gyroscopes [16].

**Acknowledgements**

Funding from the Cambridge Trusts is gratefully acknowledged.

**Authors' contributions**

AG and AAS conceived the idea; AG and CD designed the device and performed the experiments; AG analysed the results and wrote the manuscript; AAS supervised the research.

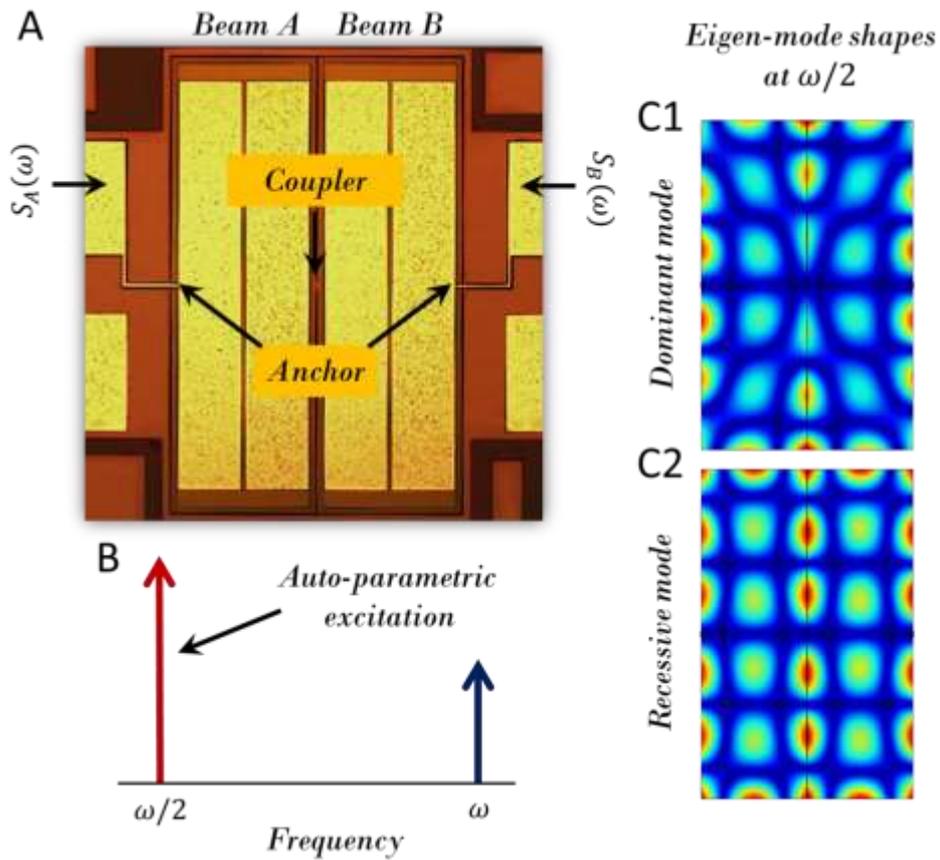

Figure 1: **Auto-parametric excitation in coupled micromechanical beams.** A: Signals $S_A(\omega)$ and $S_B(\omega)$ are applied on a mechanically coupled free-free beam microstructure; B: Auto-parametric excitation of the tone at $\omega/2$; C1-C2: The mode shapes of dominant and recessive modes at around $\omega/2 = 1.93\ MHz$ predicted using COMSOL Multiphysics.

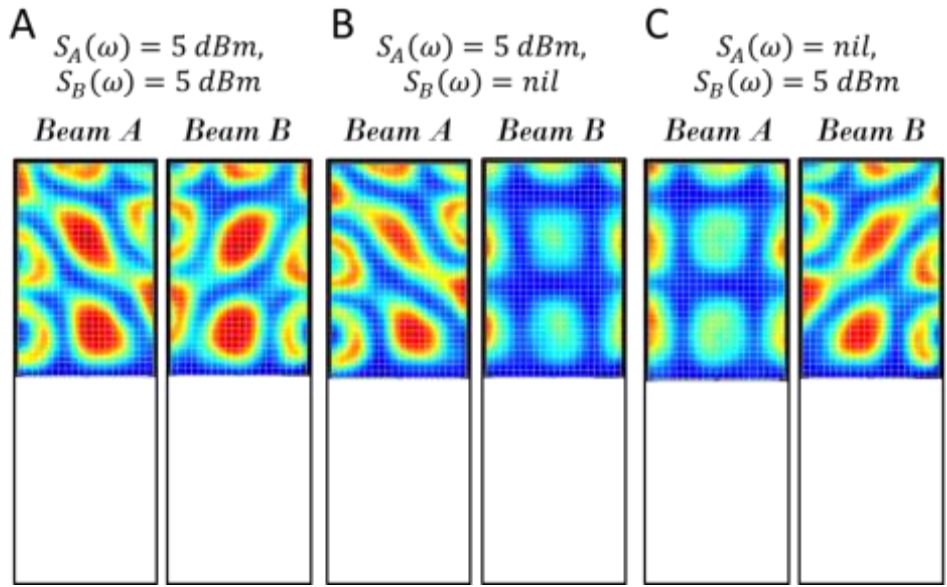

Figure 2: **Anomaly in coupled parametric resonance.** The 2-D displacement profiles of beams A and B for the drive conditions, A: $S_A(\omega = 3.86\ MHz) = 5\ dBm$ and $S_B(\omega = 3.86\ MHz) = 5\ dBm$; B: $S_A(\omega = 3.86\ MHz) = 5\ dBm$ and $S_B(\omega = 3.86\ MHz) = nil$; C: $S_A(\omega = 3.86\ MHz) = nil$ and $S_B(\omega = 3.86\ MHz) = 5\ dBm$. Note: The displacement profiles are normalized to the maximum displacement in the structure- i.e. Red corresponds to 1 and blue corresponds to 0.

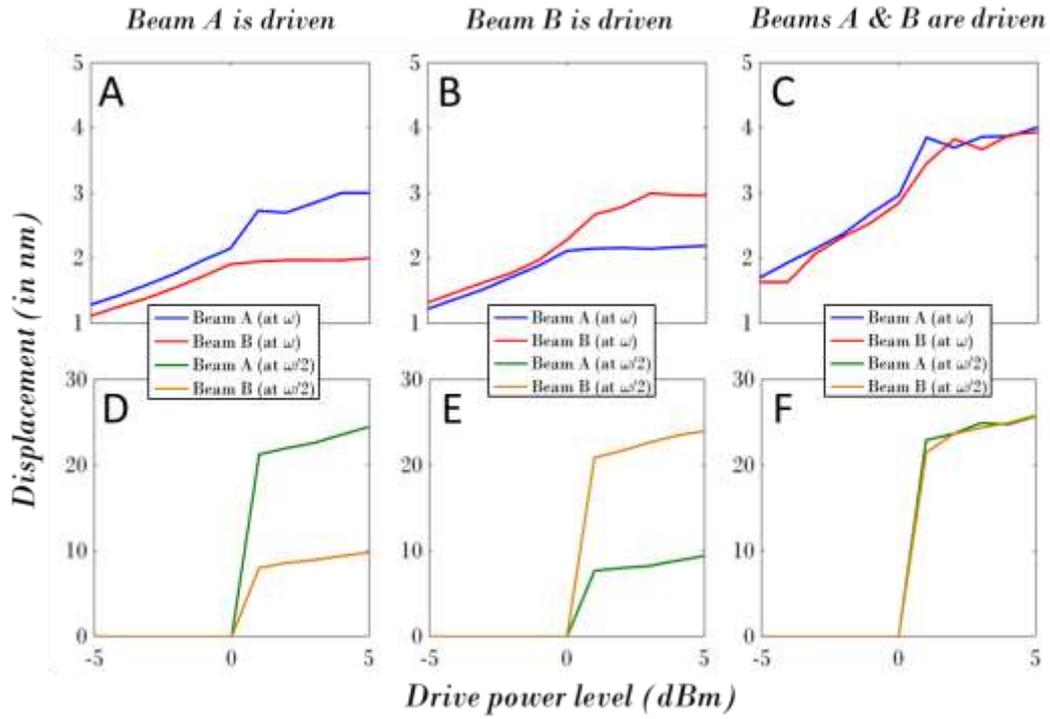

Figure 3: **Driver power level dependence.** The displacement amplitudes of the tone $\omega$ at the beams A and B for the drive conditions, A: $S_A(\omega = 3.86\ MHz) = 5\ dBm$ and $S_B(\omega = 3.86\ MHz) = nil$; B: $S_A(\omega = 3.86\ MHz) = nil$ and $S_B(\omega = 3.86\ MHz) = 5\ dBm$; C: $S_A(\omega = 3.86\ MHz) = 5\ dBm$ and $S_B(\omega = 3.86\ MHz) = 5\ dBm$. The displacement amplitudes of the tone $\omega/2$ at the beams A and B for the drive conditions, D: $S_A(\omega = 3.86\ MHz) = 5\ dBm$ and $S_B(\omega = 3.86\ MHz) = nil$; E: $S_A(\omega = 3.86\ MHz) = nil$ and $S_B(\omega = 3.86\ MHz) = 5\ dBm$; F: $S_A(\omega = 3.86\ MHz) = 5\ dBm$ and $S_B(\omega = 3.86\ MHz) = 5\ dBm$.

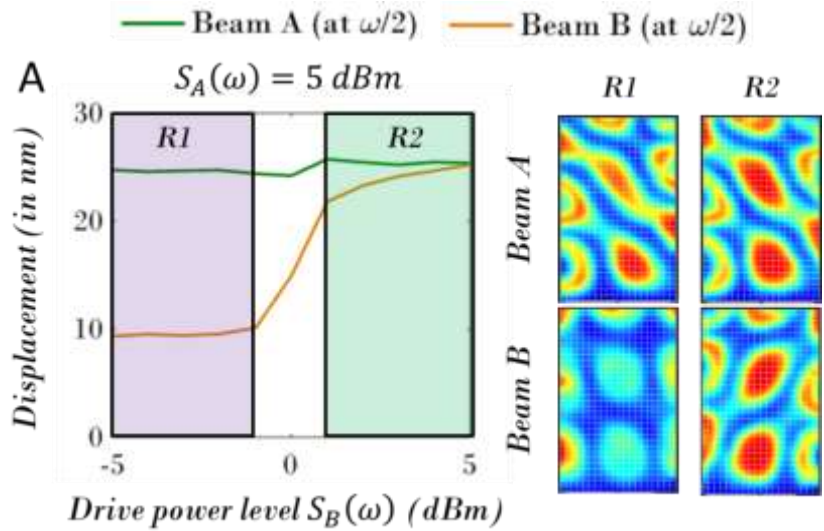

Figure 4: **Crossover from dominant to recessive mode excitation.** A: The displacement amplitudes of the tone $\omega/2$ at the beams A and B for the drive conditions $S_A(\omega = 3.86\ MHz) = 5\ dBm$ and $S_B(\omega = 3.86\ MHz) = -5\ dBm\ to\ 5\ dBm$; The 2-D displacement profiles of beams A and B for the drive conditions, R1: $S_A(\omega = 3.86\ MHz) = 5\ dBm$ and $S_B(\omega = 3.86\ MHz) = -5\ dBm$; R2: $S_A(\omega = 3.86\ MHz) = 5\ dBm$ and $S_B(\omega = 3.86\ MHz) = 5\ dBm$. Note: The displacement profiles are normalized to the maximum displacement in the structure- i.e. Red corresponds to 1 and blue corresponds to 0.

**Supplementary Information**

**Anomaly in coupled parametric resonance**

Authors: Adarsh Ganesan[1], Cuong Do[1], Ashwin Seshia[1]

1. Nanoscience Centre, University of Cambridge, Cambridge, UK

**Supplementary section S1**

Design of piezoelectrically driven micromechanical resonator

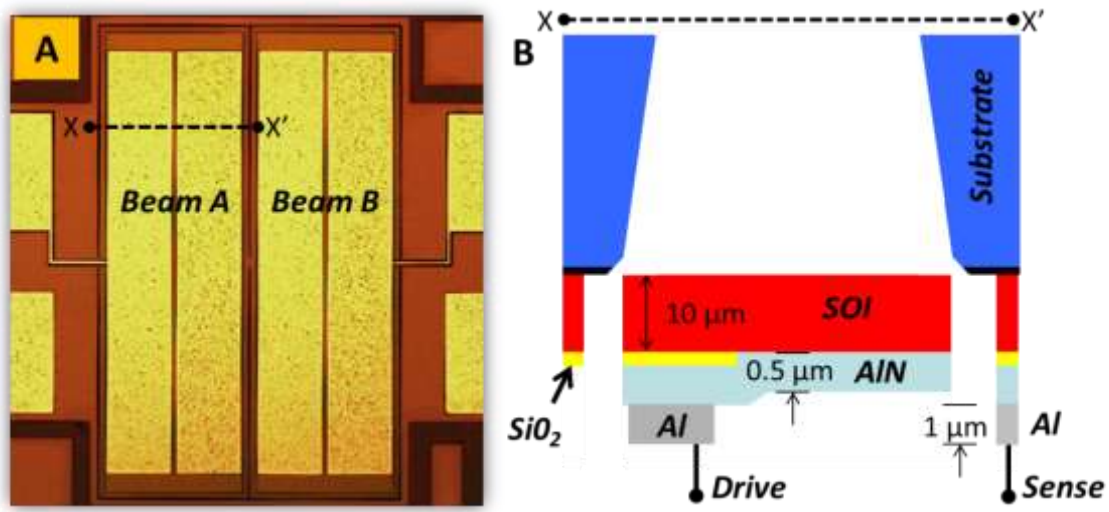

**Figure S1: Design of piezoelectrically driven micromechanical resonator**: **A**: Two coupled free-free beam topology with 2 µm air gap; **B**: 1 µm thick Al electrodes patterned on 0.5 µm thick AlN piezoelectric film which is in-turn patterned on SOI substrate; the 10 µm thick SOI layer is then released through back-side etch to realize mechanical functionality

## Supplementary section S2

Surface displacement profiles obtained using Laser Doppler Vibrometry at different drive conditions

| $S_B(\omega = 3.86\ MHz) = nil$ | | |
|---|---|---|
| Drive power level $S_A(\omega)$ (dBm) | Beam A ($\omega/2$) | Beam B ($\omega/2$) |
| -5 | - | - |
| -4 | - | - |
| -3 | - | - |
| -2 | - | - |
| -1 | - | - |
| 0 | - | - |
| 1 | 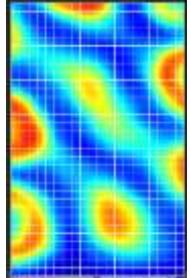 | 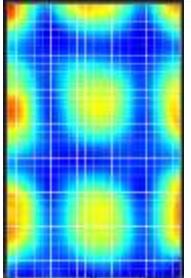 |
| 2 | 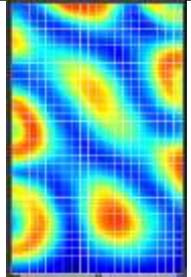 | 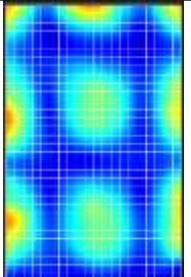 |
| 3 | 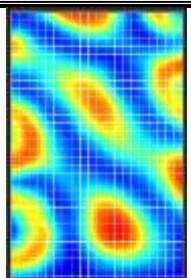 | 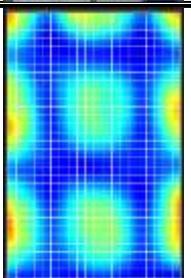 |
| 4 | 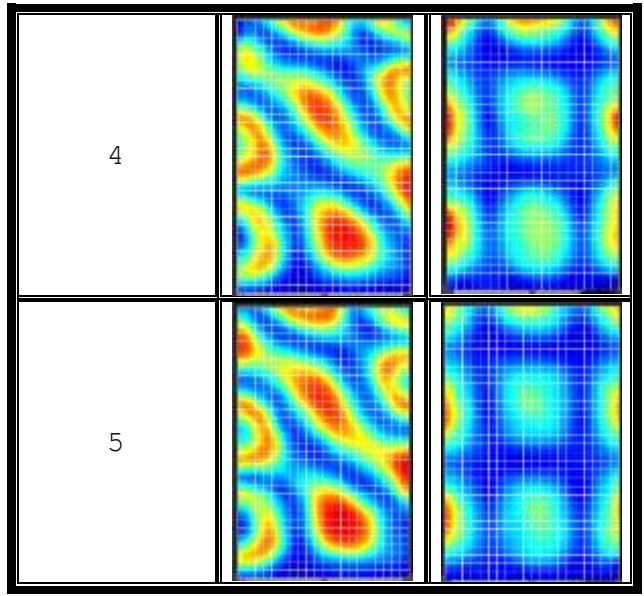 | |
| 5 | | |

| $S_A(\omega = 3.86\ MHz) = nil$ | | |
|---|---|---|
| Drive power level $S_B(\omega)$ (dBm) | Beam A ($\omega/2$) | Beam B ($\omega/2$) |
| −5 | - | - |
| −4 | - | - |
| −3 | - | - |
| −2 | - | - |
| −1 | - | - |
| 0 | - | - |
| 1 | | |

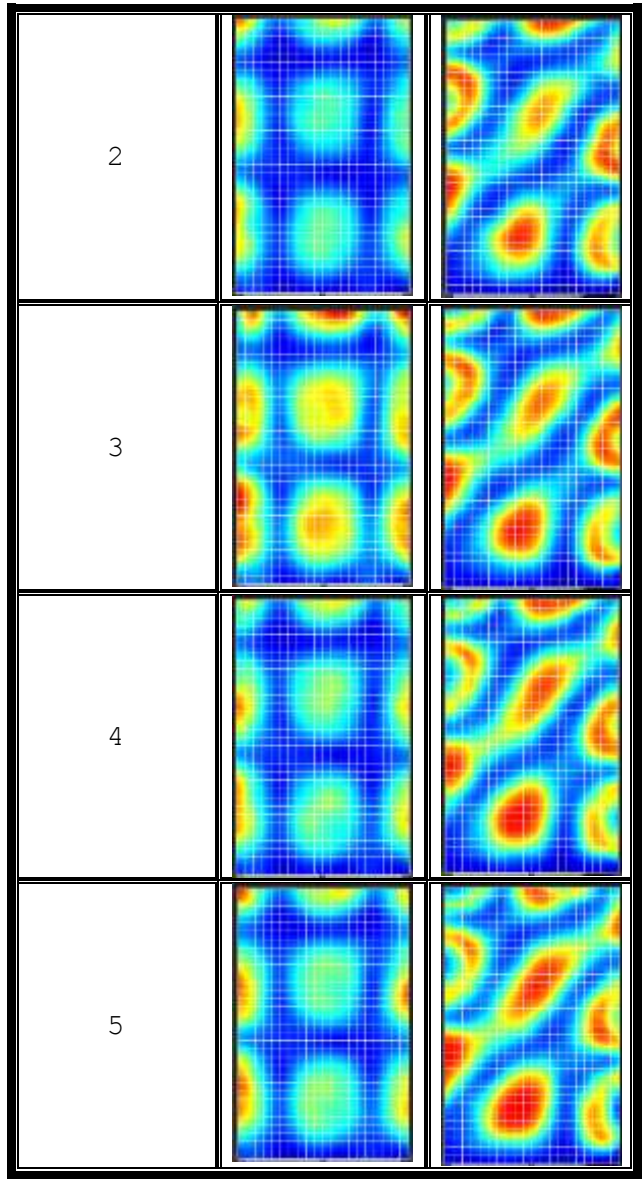

| Drive power level $S_A(\omega) = S_B(\omega)$ (dBm) | Beam A ($\omega/2$) | Beam B ($\omega/2$) |
|---|---|---|
| −5 | - | - |
| −4 | - | - |
| −3 | - | - |
| −2 | - | - |
| −1 | - | - |

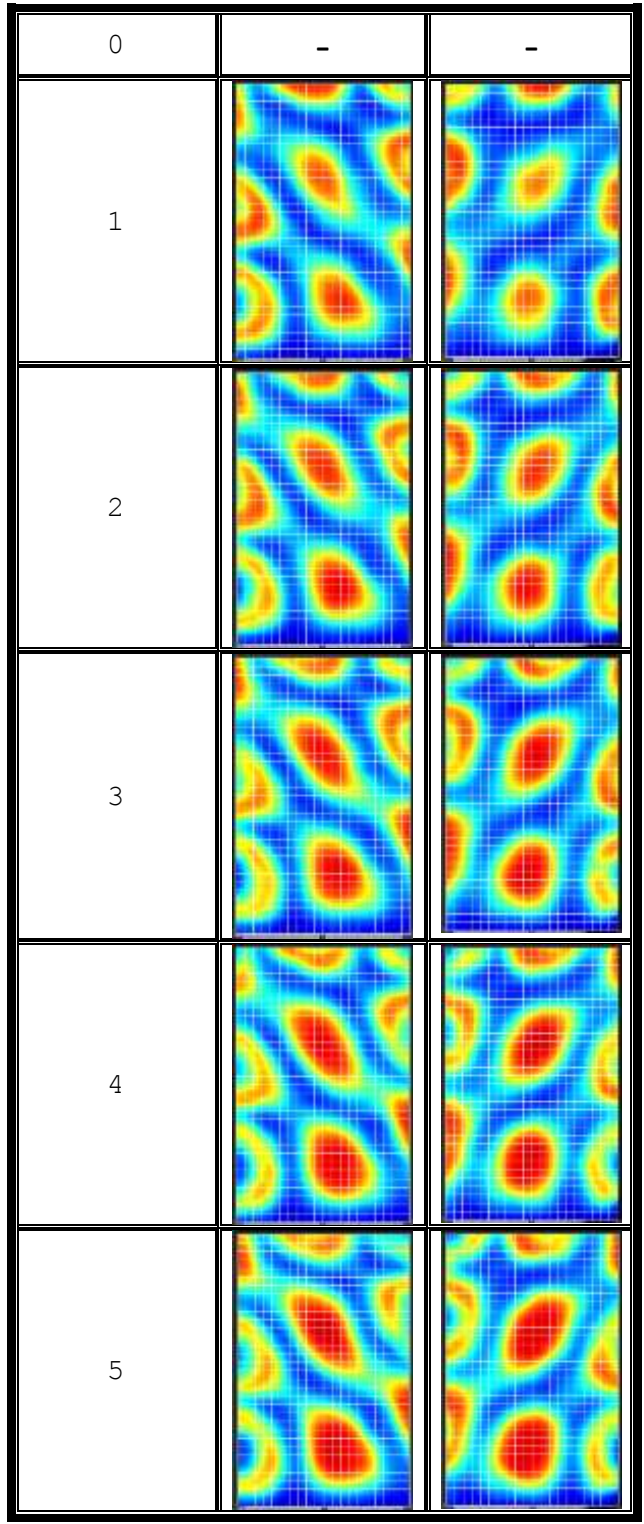

| Drive power level $S_B(\omega)$ (dBm) | Beam A ($\omega/2$) | Beam B ($\omega/2$) |
|---|---|---|
| 0 | - | - |
| 1 | | |
| 2 | | |
| 3 | | |
| 4 | | |
| 5 | | |

$S_A(\omega = 3.86\ MHz) = 5\ dBm$

| −5 | | |
|---|---|---|
| −4 | | |
| −3 | | |
| −2 | | |
| −1 | | |
| 0 | | |

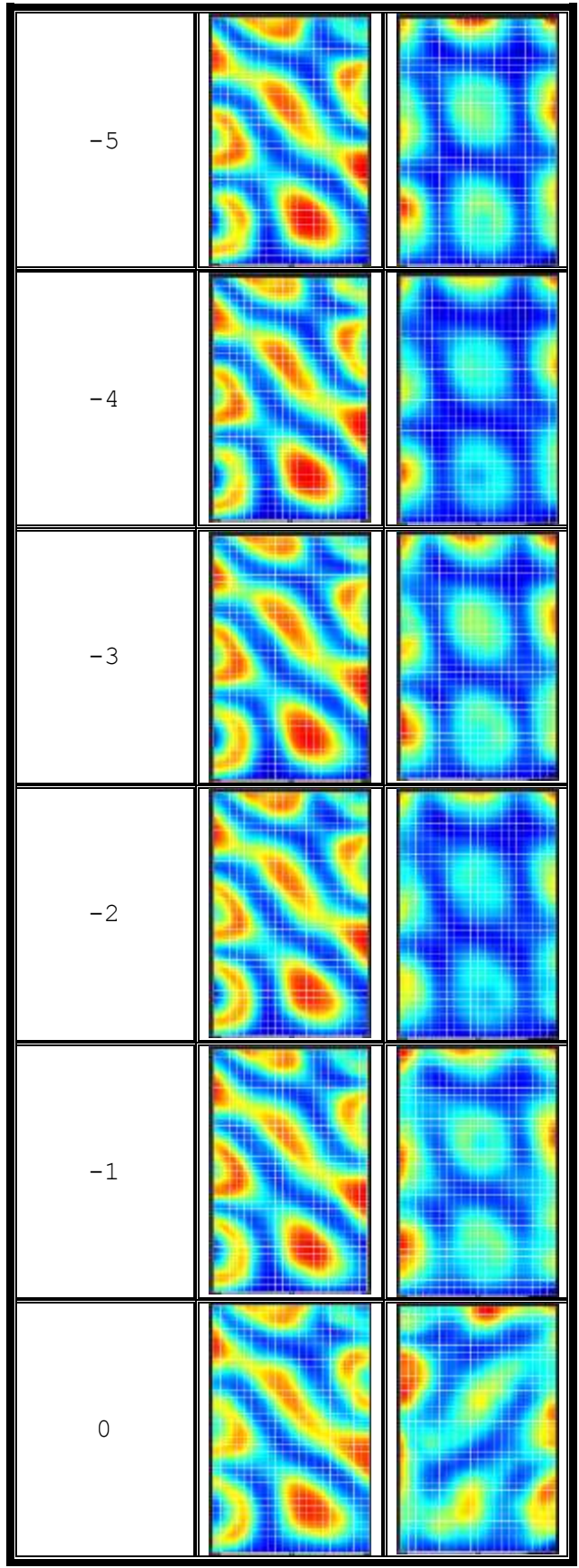

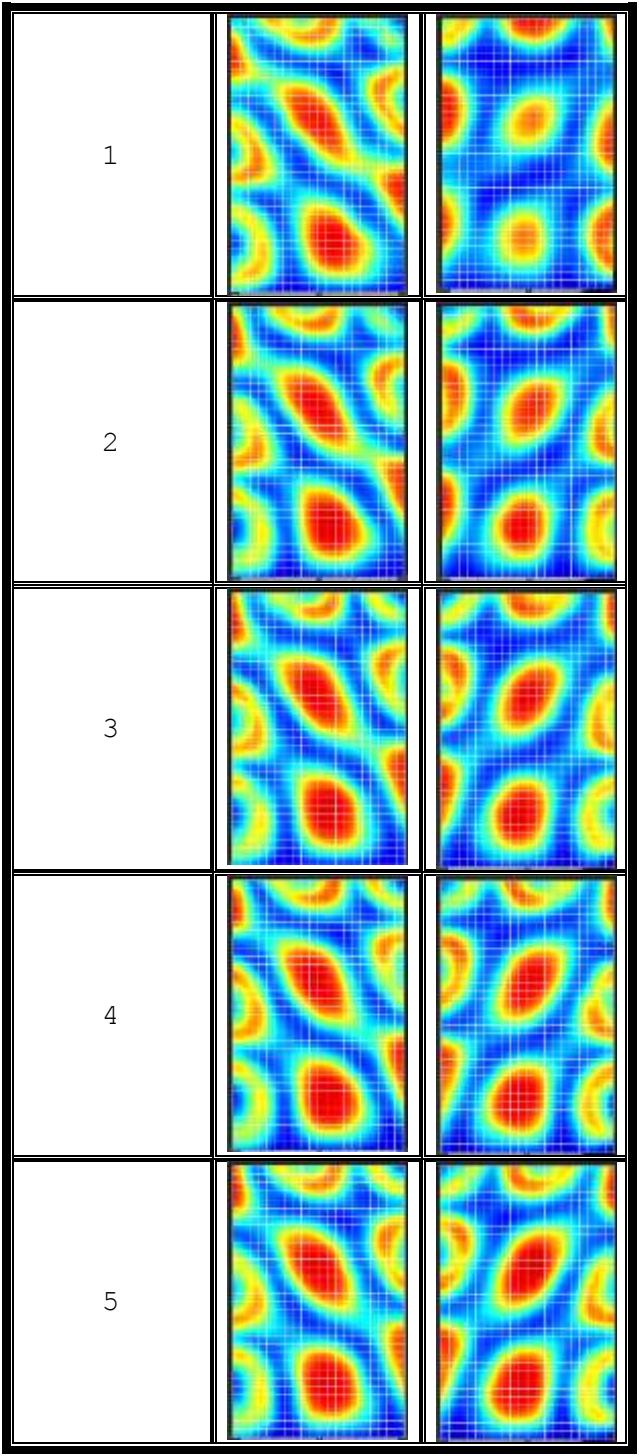